\documentclass[12pt, twoside]{article}
\usepackage{latexsym}
\usepackage{amsmath, amsfonts, amscd, amssymb}
\usepackage{amsthm}
\usepackage{fancybox}
\usepackage{eufrak, euscript, oldgerm, graphics}
\usepackage{pst-all}        
\usepackage{pst-poly}     
\usepackage{multido}      
\newcommand{\aut}{{\mathcal{A}}ut}

\newcommand{\conn}{\mathcal{D}}

\newcommand{\field}{\mathfrak{F}}

\newcommand{\ricci}{\EuScript{R}}

\newcommand{\modl}{\mathbf{\mathcal{E}}}

\newcommand{\smooth}{\mathcal{C}^{\infty}}
\newcommand{\sconn}{\textsf{A}}

\newcommand{\struc}{\mathbf{A}}

\newcommand{\bull}{{\scriptstyle\bullet}}

\newcommand{\com}{\mathbb{C}}
\newcommand{\mapto}{\longrightarrow}

\newcommand{\R}{\mathbb{R}}

\makeatletter

\def\diagram{\m@th\leftwidth=\z@ \rightwidth=\z@ \topheight=\z@
\botheight=\z@ \setbox\@picbox\hbox\bgroup}

\def\enddiagram{\egroup\wd\@picbox\rightwidth\unitlength
\ht\@picbox\topheight\unitlength \dp\@picbox\botheight\unitlength
\hskip\leftwidth\unitlength\box\@picbox}

\def\bfig{\begin{diagram}}
\def\efig{\end{diagram}}
\newcount\wideness \newcount\leftwidth \newcount\rightwidth
\newcount\highness \newcount\topheight \newcount\botheight

\def\ratchet#1#2{\ifnum#1<#2 \global #1=#2 \fi}

\def\putbox(#1,#2)#3{%
\horsize{\wideness}{#3} \divide\wideness by 2 {\advance\wideness
by #1 \ratchet{\rightwidth}{\wideness}} {\advance\wideness by -#1
\ratchet{\leftwidth}{\wideness}} \vertsize{\highness}{#3}
\divide\highness by 2 {\advance\highness by #2
\ratchet{\topheight}{\highness}} {\advance\highness by -#2
\ratchet{\botheight}{\highness}} \put(#1,#2){\makebox(0,0){$#3$}}}

\def\putlbox(#1,#2)#3{%
\horsize{\wideness}{#3} {\advance\wideness by #1
\ratchet{\rightwidth}{\wideness}} {\ratchet{\leftwidth}{-#1}}
\vertsize{\highness}{#3} \divide\highness by 2 {\advance\highness
by #2 \ratchet{\topheight}{\highness}} {\advance\highness by -#2
\ratchet{\botheight}{\highness}}
\put(#1,#2){\makebox(0,0)[l]{$#3$}}}

\def\putrbox(#1,#2)#3{%
\horsize{\wideness}{#3} {\ratchet{\rightwidth}{#1}}
{\advance\wideness by -#1 \ratchet{\leftwidth}{\wideness}}
\vertsize{\highness}{#3} \divide\highness by 2 {\advance\highness
by #2 \ratchet{\topheight}{\highness}} {\advance\highness by -#2
\ratchet{\botheight}{\highness}}
\put(#1,#2){\makebox(0,0)[r]{$#3$}}}

\def\adjust[#1]{} 

\newcount \coefa
\newcount \coefb
\newcount \coefc
\newcount\tempcounta
\newcount\tempcountb
\newcount\tempcountc
\newcount\tempcountd
\newcount\xext
\newcount\yext
\newcount\xoff
\newcount\yoff
\newcount\gap%
\newcount\arrowtypea
\newcount\arrowtypeb
\newcount\arrowtypec
\newcount\arrowtyped
\newcount\arrowtypee
\newcount\height
\newcount\width
\newcount\xpos
\newcount\ypos
\newcount\run
\newcount\rise
\newcount\arrowlength
\newcount\halflength
\newcount\arrowtype
\newdimen\tempdimen
\newdimen\xlen
\newdimen\ylen
\newsavebox{\tempboxa}%
\newsavebox{\tempboxb}%
\newsavebox{\tempboxc}%

\newdimen\w@dth

\def\setw@dth#1#2{\setbox\z@\hbox{\m@th$#1$}\w@dth=\wd\z@
\setbox\@ne\hbox{\m@th$#2$}\ifnum\w@dth<\wd\@ne \w@dth=\wd\@ne \fi
\advance\w@dth by 1.2em}

\def\t@^#1_#2{\allowbreak\def\n@one{#1}\def\n@two{#2}\mathrel
{\setw@dth{#1}{#2} \mathop{\hbox to
\w@dth{\rightarrowfill}}\limits \ifx\n@one\empty\else
^{\box\z@}\fi \ifx\n@two\empty\else _{\box\@ne}\fi}}
\def\t@@^#1{\@ifnextchar_{\t@^{#1}}{\t@^{#1}_{}}}
\def\to{\@ifnextchar^{\t@@}{\t@@^{}}}

\def\t@left^#1_#2{\def\n@one{#1}\def\n@two{#2}\mathrel{\setw@dth{#1}{#2}
\mathop{\hbox to \w@dth{\leftarrowfill}}\limits
\ifx\n@one\empty\else ^{\box\z@}\fi \ifx\n@two\empty\else
_{\box\@ne}\fi}}
\def\t@@left^#1{\@ifnextchar_{\t@left^{#1}}{\t@left^{#1}_{}}}
\def\toleft{\@ifnextchar^{\t@@left}{\t@@left^{}}}

\def\two@^#1_#2{\allowbreak
\def\n@one{#1}\def\n@two{#2}\mathrel{\setw@dth{#1}{#2}
\mathop{\vcenter{\lineskip\z@\baselineskip\z@
                 \hbox to \w@dth{\rightarrowfill}%
                 \hbox to \w@dth{\rightarrowfill}}%
       }\limits
\ifx\n@one\empty\else ^{\box\z@}\fi \ifx\n@two\empty\else
_{\box\@ne}\fi}}
\def\tw@@^#1{\@ifnextchar _{\two@^{#1}}{\two@^{#1}_{}}}
\def\two{\@ifnextchar ^{\tw@@}{\tw@@^{}}}

\def\tofr@^#1_#2{\def\n@one{#1}\def\n@two{#2}\mathrel{\setw@dth{#1}{#2}
\mathop{\vcenter{\hbox to \w@dth{\rightarrowfill}\kern-1.7ex
                 \hbox to \w@dth{\leftarrowfill}}%
       }\limits
\ifx\n@one\empty\else ^{\box\z@}\fi \ifx\n@two\empty\else
_{\box\@ne}\fi}}
\def\t@fr@^#1{\@ifnextchar_ {\tofr@^{#1}}{\tofr@^{#1}_{}}}
\def\tofro{\@ifnextchar^ {\t@fr@}{\t@fr@^{}}}

\def\mon{\mathop{\m@th\hbox to
      14.6\P@{\lasyb\char'51\hskip-2.1\P@$\arrext$\hss
$\mathord\rightarrow$}}\limits} 
\def\leftmono{\mathrel{\m@th\hbox to
14.6\P@{$\mathord\leftarrow$\hss$\arrext$\hskip-2.1\P@\lasyb\char'50%
}}\limits} 
\mathchardef\arrext="0200       

\def\settypes(#1,#2,#3){\arrowtypea#1 \arrowtypeb#2 \arrowtypec#3}
\def\settoheight#1#2{\setbox\@tempboxa\hbox{#2}#1\ht\@tempboxa\relax}%
\def\settodepth#1#2{\setbox\@tempboxa\hbox{#2}#1\dp\@tempboxa\relax}%
\def\settokens`#1`#2`#3`#4`{%
     \def\tokena{#1}\def\tokenb{#2}\def\tokenc{#3}\def\tokend{#4}}
\def\setsqparms[#1`#2`#3`#4;#5`#6]{%
\arrowtypea #1 \arrowtypeb #2 \arrowtypec #3 \arrowtyped #4 \width
#5 \height #6 }
\def\setpos(#1,#2){\xpos=#1 \ypos#2}

\def\settriparms[#1`#2`#3;#4]{\settripairparms[#1`#2`#3`1`1;#4]}%

\def\settripairparms[#1`#2`#3`#4`#5;#6]{%
\arrowtypea #1 \arrowtypeb #2 \arrowtypec #3 \arrowtyped #4
\arrowtypee #5 \width #6 \height #6 }

\def\resetparms{\settripairparms[1`1`1`1`1;500]\width 500}

\resetparms

\def\mvector(#1,#2)#3{
\put(0,0){\vector(#1,#2){#3}}%
\put(0,0){\vector(#1,#2){26}}%
}
\def\evector(#1,#2)#3{{
\arrowlength #3
\put(0,0){\vector(#1,#2){\arrowlength}}%
\advance \arrowlength by-30
\put(0,0){\vector(#1,#2){\arrowlength}}%
}}

\def\horsize#1#2{%
\settowidth{\tempdimen}{$#2$}%
#1=\tempdimen \divide #1 by\unitlength }

\def\vertsize#1#2{%
\settoheight{\tempdimen}{$#2$}%
#1=\tempdimen
\settodepth{\tempdimen}{$#2$}%
\advance #1 by\tempdimen \divide #1 by\unitlength }

\def\putvector(#1,#2)(#3,#4)#5#6{{%
\ifnum3<\arrowtype \putdashvector(#1,#2)(#3,#4)#5\arrowtype \else
\ifnum\arrowtype<-3 \putdashvector(#1,#2)(#3,#4)#5\arrowtype \else
\xpos=#1 \ypos=#2 \run=#3 \rise=#4 \arrowlength=#5 \ifnum
\arrowtype<0
    \ifnum \run=0
        \advance \ypos by-\arrowlength
    \else
        \tempcounta \arrowlength
        \multiply \tempcounta by\rise
        \divide \tempcounta by\run
        \ifnum\run>0
            \advance \xpos by\arrowlength
            \advance \ypos by\tempcounta
        \else
            \advance \xpos by-\arrowlength
            \advance \ypos by-\tempcounta
        \fi
    \fi
    \multiply \arrowtype by-1
    \multiply \rise by-1
    \multiply \run by-1
\fi \ifcase \arrowtype
\or \put(\xpos,\ypos){\vector(\run,\rise){\arrowlength}}%
\or \put(\xpos,\ypos){\mvector(\run,\rise)\arrowlength}%
\or \put(\xpos,\ypos){\evector(\run,\rise){\arrowlength}}%
\fi\fi\fi }}

\def\putsplitvector(#1,#2)#3#4{
\xpos #1 \ypos #2 \arrowtype #4 \halflength #3 \arrowlength #3
\gap 140 \advance \halflength by-\gap \divide \halflength by2
\ifnum\arrowtype>0
   \ifcase \arrowtype
   \or \put(\xpos,\ypos){\line(0,-1){\halflength}}%
       \advance\ypos by-\halflength
       \advance\ypos by-\gap
       \put(\xpos,\ypos){\vector(0,-1){\halflength}}%
   \or \put(\xpos,\ypos){\line(0,-1)\halflength}%
       \put(\xpos,\ypos){\vector(0,-1)3}%
       \advance\ypos by-\halflength
       \advance\ypos by-\gap
       \put(\xpos,\ypos){\vector(0,-1){\halflength}}%
   \or \put(\xpos,\ypos){\line(0,-1)\halflength}%
       \advance\ypos by-\halflength
       \advance\ypos by-\gap
       \put(\xpos,\ypos){\evector(0,-1){\halflength}}%
   \fi
\else \arrowtype=-\arrowtype
   \ifcase\arrowtype
   \or \advance \ypos by-\arrowlength
       \put(\xpos,\ypos){\line(0,1){\halflength}}%
       \advance\ypos by\halflength
       \advance\ypos by\gap
       \put(\xpos,\ypos){\vector(0,1){\halflength}}%
   \or \advance \ypos by-\arrowlength
       \put(\xpos,\ypos){\line(0,1)\halflength}%
       \put(\xpos,\ypos){\vector(0,1)3}%
       \advance\ypos by\halflength
       \advance\ypos by\gap
       \put(\xpos,\ypos){\vector(0,1){\halflength}}%
   \or \advance \ypos by-\arrowlength
       \put(\xpos,\ypos){\line(0,1)\halflength}%
       \advance\ypos by\halflength
       \advance\ypos by\gap
       \put(\xpos,\ypos){\evector(0,1){\halflength}}%
   \fi
\fi }

\def\putmorphism(#1)(#2,#3)[#4`#5`#6]#7#8#9{{%
\run #2 \rise #3 \ifnum\rise=0
  \puthmorphism(#1)[#4`#5`#6]{#7}{#8}#9%
\else\ifnum\run=0
  \putvmorphism(#1)[#4`#5`#6]{#7}{#8}#9%
\else
\setpos(#1)%
\arrowlength #7 \arrowtype #8 \ifnum\run=0 \else\ifnum\rise=0
\else \ifnum\run>0
    \coefa=1
\else
   \coefa=-1
\fi \ifnum\arrowtype>0
   \coefb=0
   \coefc=-1
\else
   \coefb=\coefa
   \coefc=1
   \arrowtype=-\arrowtype
\fi \width=2 \multiply \width by\run \divide \width by\rise \ifnum
\width<0  \width=-\width\fi \advance\width by60 \if l#9
\width=-\width\fi
\putbox(\xpos,\ypos){#4}
{\multiply \coefa by\arrowlength
\advance\xpos by\coefa \multiply \coefa by\rise \divide \coefa
by\run \advance \ypos by\coefa
\putbox(\xpos,\ypos){#5} }%
{\multiply \coefa by\arrowlength
\divide \coefa by2 \advance \xpos by\coefa \advance \xpos by\width
\multiply \coefa by\rise \divide \coefa by\run \advance \ypos
by\coefa
\if l#9%
   \putrbox(\xpos,\ypos){#6}%
\else\if r#9%
   \putlbox(\xpos,\ypos){#6}%
\fi\fi }%
{\multiply \rise by-\coefc
\multiply \run by-\coefc \multiply \coefb by\arrowlength \advance
\xpos by\coefb \multiply \coefb by\rise \divide \coefb by\run
\advance \ypos by\coefb \multiply \coefc by70 \advance \ypos
by\coefc \multiply \coefc by\run \divide \coefc by\rise \advance
\xpos by\coefc \multiply \coefa by140 \multiply \coefa by\run
\divide \coefa by\rise \advance \arrowlength by\coefa
\ifcase\arrowtype
\or \put(\xpos,\ypos){\vector(\run,\rise){\arrowlength}}%
\or \put(\xpos,\ypos){\mvector(\run,\rise){\arrowlength}}%
\or \put(\xpos,\ypos){\evector(\run,\rise){\arrowlength}}%
\fi}\fi\fi\fi\fi}}

\newcount\numbdashes \newcount\lengthdash \newcount\increment

\def\howmanydashes{
\numbdashes=\arrowlength \lengthdash=40 \divide\numbdashes by
\lengthdash \lengthdash=\arrowlength \divide\lengthdash by
\numbdashes
\increment=\lengthdash \multiply\lengthdash by 3
\divide\lengthdash by 5 }

\def\putdashvector(#1)(#2,#3)#4#5{%
\ifnum#3=0 \putdashhvector(#1){#4}#5 \else \ifnum#2=0
\putdashvvector(#1){#4}#5\fi\fi}

\def\putdashhvector(#1,#2)#3#4{{%
\arrowlength=#3 \howmanydashes
\multiput(#1,#2)(\increment,0){\numbdashes}%
{\vrule height .4pt width \lengthdash\unitlength} \arrowtype=#4
\xpos=#1 \ifnum\arrowtype<0 \advance\arrowtype by 7 \fi
\ifcase\arrowtype \or \advance\xpos by 10
    \put(\xpos,#2){\vector(-1,0){\lengthdash}}
    \advance\xpos by 40
    \put(\xpos,#2){\vector(-1,0){\lengthdash}}
\or \advance \xpos by 10
    \put(\xpos,#2){\vector(-1,0){\lengthdash}}
    \advance\xpos by  \arrowlength
    \advance\xpos by  -50
    \put(\xpos,#2){\vector(-1,0){\lengthdash}}
\or \advance\xpos by 10
    \put(\xpos,#2){\vector(-1,0){\lengthdash}}
\or \advance\xpos by \arrowlength
    \advance\xpos by -\lengthdash
    \put(\xpos,#2){\vector(1,0){\lengthdash}}
\or {\advance\xpos by 10
    \put(\xpos,#2){\vector(1,0){\lengthdash}}}
    \advance\xpos by \arrowlength
    \advance\xpos by -\lengthdash
    \put(\xpos,#2){\vector(1,0){\lengthdash}}
\or \advance\xpos by \arrowlength
    \advance\xpos by -\lengthdash
    \put(\xpos,#2){\vector(1,0){\lengthdash}}
    \advance\xpos by -40
    \put(\xpos,#2){\vector(1,0){\lengthdash}}
   \fi
}}

\def\putdashvvector(#1,#2)#3#4{{%
\arrowlength=#3 \howmanydashes \ypos=#2 \advance\ypos by
-\arrowlength
\multiput(#1,#2)(0,\increment){\numbdashes}%
    {\vrule width .4pt height \lengthdash\unitlength}
\arrowtype=#4 \ypos=#2 \ifnum\arrowtype<0 \advance\arrowtype by 7
\fi \ifcase\arrowtype \or \advance\ypos by \arrowlength
\advance\ypos by -40
    \put(#1,\ypos){\vector(0,1){\lengthdash}}
    \advance\ypos by -40
    \put(#1,\ypos){\vector(0,1){\lengthdash}}
\or \advance\ypos by 10
    \put(#1,\ypos){\vector(0,1){\lengthdash}}
    \advance\ypos by \arrowlength \advance\ypos by -40
    \put(#1,\ypos){\vector(0,1){\lengthdash}}
\or \advance\ypos by \arrowlength \advance\ypos by -40
    \put(#1,\ypos){\vector(0,1){\lengthdash}}
\or \advance\ypos by 10
    \put(#1,\ypos){\vector(0,-1){\lengthdash}}
\or \advance\ypos by 10
    \put(#1,\ypos){\vector(0,-1){\lengthdash}}
    \advance\ypos by \arrowlength \advance\ypos by -40
    \put(#1,\ypos){\vector(0,-1){\lengthdash}}
\or \advance\ypos by 10
    \put(#1,\ypos){\vector(0,-1){\lengthdash}}
    \advance\ypos by 40
    \put(#1,\ypos){\vector(0,-1){\lengthdash}}
\fi }}

\def\puthmorphism(#1,#2)[#3`#4`#5]#6#7#8{{%
\xpos #1 \ypos #2 \width #6 \arrowlength #6 \arrowtype=#7
\putbox(\xpos,\ypos){#3\vphantom{#4}}%
{\advance \xpos by\arrowlength
\putbox(\xpos,\ypos){\vphantom{#3}#4}}%
\horsize{\tempcounta}{#3}%
\horsize{\tempcountb}{#4}%
\divide \tempcounta by2 \divide \tempcountb by2 \advance
\tempcounta by30 \advance \tempcountb by30 \advance \xpos
by\tempcounta \advance \arrowlength by-\tempcounta \advance
\arrowlength by-\tempcountb
\putvector(\xpos,\ypos)(1,0)\arrowlength\arrowtype \divide
\arrowlength by2 \advance \xpos by\arrowlength
\vertsize{\tempcounta}{#5}%
\divide\tempcounta by2 \advance \tempcounta by20
\if a#8 %
   \advance \ypos by\tempcounta
   \putbox(\xpos,\ypos){#5}%
\else
   \advance \ypos by-\tempcounta
   \putbox(\xpos,\ypos){#5}%
\fi}}

\def\putvmorphism(#1,#2)[#3`#4`#5]#6#7#8{{%
\xpos #1 \ypos #2 \arrowlength #6 \arrowtype #7
\settowidth{\xlen}{$#5$}%
\putbox(\xpos,\ypos){#3}%
{\advance \ypos by-\arrowlength
\putbox(\xpos,\ypos){#4}}%
{\advance\arrowlength by-140 \advance \ypos by-70 \ifdim\xlen>0pt
   \if m#8%
      \putsplitvector(\xpos,\ypos)\arrowlength\arrowtype
   \else
   \putvector(\xpos,\ypos)(0,-1)\arrowlength\arrowtype
   \fi
\else
   \putvector(\xpos,\ypos)(0,-1)\arrowlength\arrowtype
\fi}%
\ifdim\xlen>0pt
   \divide \arrowlength by2
   \advance\ypos by-\arrowlength
   \if l#8%
      \advance \xpos by-40
      \putrbox(\xpos,\ypos){#5}%
   \else\if r#8%
      \advance \xpos by40
      \putlbox(\xpos,\ypos){#5}%
   \else
      \putbox(\xpos,\ypos){#5}%
   \fi\fi
\fi }}

\def\putsquarep<#1>(#2)[#3;#4`#5`#6`#7]{{%
\setsqparms[#1]%
\setpos(#2)%
\settokens`#3`%
\puthmorphism(\xpos,\ypos)[\tokenc`\tokend`{#7}]{\width}{\arrowtyped}b%
\advance\ypos by \height
\puthmorphism(\xpos,\ypos)[\tokena`\tokenb`{#4}]{\width}{\arrowtypea}a%
\putvmorphism(\xpos,\ypos)[``{#5}]{\height}{\arrowtypeb}l%
\advance\xpos by \width
\putvmorphism(\xpos,\ypos)[``{#6}]{\height}{\arrowtypec}r%
}}

\def\putsquare{\@ifnextchar <{\putsquarep}{\putsquarep%
   <\arrowtypea`\arrowtypeb`\arrowtypec`\arrowtyped;\width`\height>}}
\def\square{\@ifnextchar< {\squarep}{\squarep
   <\arrowtypea`\arrowtypeb`\arrowtypec`\arrowtyped;\width`\height>}}
\def\squarep<#1>[#2`#3`#4`#5;#6`#7`#8`#9]{{
\setsqparms[#1]
\diagram
\putsquarep<\arrowtypea`\arrowtypeb`\arrowtypec`
\arrowtyped;\width`\height>
(0,0)[#2`#3`#4`{#5};#6`#7`#8`{#9}]
\enddiagram
}}                                                 
\def\putptrianglep<#1>(#2,#3)[#4`#5`#6;#7`#8`#9]{{%
\settriparms[#1]%
\xpos=#2 \ypos=#3 \advance\ypos by \height
\puthmorphism(\xpos,\ypos)[#4`#5`{#7}]{\height}{\arrowtypea}a%
\putvmorphism(\xpos,\ypos)[`#6`{#8}]{\height}{\arrowtypeb}l%
\advance\xpos by\height
\putmorphism(\xpos,\ypos)(-1,-1)[``{#9}]{\height}{\arrowtypec}r%
}}

\def\putptriangle{\@ifnextchar <{\putptrianglep}{\putptrianglep
   <\arrowtypea`\arrowtypeb`\arrowtypec;\height>}}
\def\ptriangle{\@ifnextchar <{\ptrianglep}{\ptrianglep
   <\arrowtypea`\arrowtypeb`\arrowtypec;\height>}}
\def\ptrianglep<#1>[#2`#3`#4;#5`#6`#7]{{
\settriparms[#1]
\diagram
\putptrianglep<\arrowtypea`\arrowtypeb`
\arrowtypec;\height>
(0,0)[#2`#3`#4;#5`#6`{#7}]
\enddiagram
}}                                            

\def\putqtrianglep<#1>(#2,#3)[#4`#5`#6;#7`#8`#9]{{%
\settriparms[#1]%
\xpos=#2 \ypos=#3 \advance\ypos by\height
\puthmorphism(\xpos,\ypos)[#4`#5`{#7}]{\height}{\arrowtypea}a%
\putmorphism(\xpos,\ypos)(1,-1)[``{#8}]{\height}{\arrowtypeb}l%
\advance\xpos by\height
\putvmorphism(\xpos,\ypos)[`#6`{#9}]{\height}{\arrowtypec}r%
}}

\def\putqtriangle{\@ifnextchar <{\putqtrianglep}{\putqtrianglep
   <\arrowtypea`\arrowtypeb`\arrowtypec;\height>}}
\def\qtriangle{\@ifnextchar <{\qtrianglep}{\qtrianglep
   <\arrowtypea`\arrowtypeb`\arrowtypec;\height>}}
\def\qtrianglep<#1>[#2`#3`#4;#5`#6`#7]{{
\settriparms[#1]
\width=\height                                
\diagram
\putqtrianglep<\arrowtypea`\arrowtypeb`
\arrowtypec;\height>
(0,0)[#2`#3`#4;#5`#6`{#7}]
\enddiagram
}}

\def\putdtrianglep<#1>(#2,#3)[#4`#5`#6;#7`#8`#9]{{%
\settriparms[#1]%
\xpos=#2 \ypos=#3
\puthmorphism(\xpos,\ypos)[#5`#6`{#9}]{\height}{\arrowtypec}b%
\advance\xpos by \height \advance\ypos by\height
\putmorphism(\xpos,\ypos)(-1,-1)[``{#7}]{\height}{\arrowtypea}l%
\putvmorphism(\xpos,\ypos)[#4``{#8}]{\height}{\arrowtypeb}r%
}}

\def\putdtriangle{\@ifnextchar <{\putdtrianglep}{\putdtrianglep
   <\arrowtypea`\arrowtypeb`\arrowtypec;\height>}}
\def\dtriangle{\@ifnextchar <{\dtrianglep}{\dtrianglep
   <\arrowtypea`\arrowtypeb`\arrowtypec;\height>}}
\def\dtrianglep<#1>[#2`#3`#4;#5`#6`#7]{{
\settriparms[#1]
\width=\height                                
\diagram
\putdtrianglep<\arrowtypea`\arrowtypeb`
\arrowtypec;\height>
(0,0)[#2`#3`#4;#5`#6`{#7}]
\enddiagram
}}

\def\putbtrianglep<#1>(#2,#3)[#4`#5`#6;#7`#8`#9]{{%
\settriparms[#1]%
\xpos=#2 \ypos=#3
\puthmorphism(\xpos,\ypos)[#5`#6`{#9}]{\height}{\arrowtypec}b%
\advance\ypos by\height
\putmorphism(\xpos,\ypos)(1,-1)[``{#8}]{\height}{\arrowtypeb}r%
\putvmorphism(\xpos,\ypos)[#4``{#7}]{\height}{\arrowtypea}l%
}}

\def\putbtriangle{\@ifnextchar <{\putbtrianglep}{\putbtrianglep
   <\arrowtypea`\arrowtypeb`\arrowtypec;\height>}}
\def\btriangle{\@ifnextchar <{\btrianglep}{\btrianglep
   <\arrowtypea`\arrowtypeb`\arrowtypec;\height>}}
\def\btrianglep<#1>[#2`#3`#4;#5`#6`#7]{{
\settriparms[#1]
\width=\height                               
\diagram
\putbtrianglep<\arrowtypea`\arrowtypeb`
\arrowtypec;\height>
(0,0)[#2`#3`#4;#5`#6`{#7}]
\enddiagram
}}

\def\putAtrianglep<#1>(#2,#3)[#4`#5`#6;#7`#8`#9]{{%
\settriparms[#1]%
\xpos=#2 \ypos=#3 {\multiply \height by2
\puthmorphism(\xpos,\ypos)[#5`#6`{#9}]{\height}{\arrowtypec}b}%
\advance\xpos by\height \advance\ypos by\height
\putmorphism(\xpos,\ypos)(-1,-1)[#4``{#7}]{\height}{\arrowtypea}l%
\putmorphism(\xpos,\ypos)(1,-1)[``{#8}]{\height}{\arrowtypeb}r%
}}

\def\putAtriangle{\@ifnextchar <{\putAtrianglep}{\putAtrianglep
   <\arrowtypea`\arrowtypeb`\arrowtypec;\height>}}
\def\Atriangle{\@ifnextchar <{\Atrianglep}{\Atrianglep
   <\arrowtypea`\arrowtypeb`\arrowtypec;\height>}}
\def\Atrianglep<#1>[#2`#3`#4;#5`#6`#7]{{
\settriparms[#1]
\width=\height                                     
\diagram
\putAtrianglep<\arrowtypea`\arrowtypeb`
\arrowtypec;\height>
(0,0)[#2`#3`#4;#5`#6`{#7}]
\enddiagram
}}

\def\putAtrianglepairp<#1>(#2)[#3;#4`#5`#6`#7`#8]{{%
\settripairparms[#1]%
\setpos(#2)%
\settokens`#3`%
\puthmorphism(\xpos,\ypos)[\tokenb`\tokenc`{#7}]{\height}{\arrowtyped}b%
\advance\xpos by\height
\puthmorphism(\xpos,\ypos)[\phantom{\tokenc}`\tokend`{#8}]%
{\height}{\arrowtypee}b%
\advance\ypos by\height
\putmorphism(\xpos,\ypos)(-1,-1)[\tokena``{#4}]{\height}{\arrowtypea}l%
\putvmorphism(\xpos,\ypos)[``{#5}]{\height}{\arrowtypeb}m%
\putmorphism(\xpos,\ypos)(1,-1)[``{#6}]{\height}{\arrowtypec}r%
}}

\def\putAtrianglepair{\@ifnextchar <{\putAtrianglepairp}{\putAtrianglepairp%
   <\arrowtypea`\arrowtypeb`\arrowtypec`\arrowtyped`\arrowtypee;\height>}}
\def\Atrianglepair{\@ifnextchar <{\Atrianglepairp}{\Atrianglepairp%
   <\arrowtypea`\arrowtypeb`\arrowtypec`\arrowtyped`\arrowtypee;\height>}}

\def\Atrianglepairp<#1>[#2;#3`#4`#5`#6`#7]{{
\settripairparms[#1]
\settokens`#2`
\width=\height                                
\diagram
\putAtrianglepairp                            
<\arrowtypea`\arrowtypeb`\arrowtypec`
\arrowtyped`\arrowtypee;\height>
(0,0)[{#2};#3`#4`#5`#6`{#7}]
\enddiagram
}}

\def\putVtrianglep<#1>(#2,#3)[#4`#5`#6;#7`#8`#9]{{%
\settriparms[#1]%
\xpos=#2 \ypos=#3 \advance\ypos by\height {\multiply\height by2
\puthmorphism(\xpos,\ypos)[#4`#5`{#7}]{\height}{\arrowtypea}a}%
\putmorphism(\xpos,\ypos)(1,-1)[`#6`{#8}]{\height}{\arrowtypeb}l%
\advance\xpos by\height \advance\xpos by\height
\putmorphism(\xpos,\ypos)(-1,-1)[``{#9}]{\height}{\arrowtypec}r%
}}

\def\putVtriangle{\@ifnextchar <{\putVtrianglep}{\putVtrianglep
   <\arrowtypea`\arrowtypeb`\arrowtypec;\height>}}
\def\Vtriangle{\@ifnextchar <{\Vtrianglep}{\Vtrianglep
   <\arrowtypea`\arrowtypeb`\arrowtypec;\height>}}
\def\Vtrianglep<#1>[#2`#3`#4;#5`#6`#7]{{
\settriparms[#1]
\width=\height                                 
\diagram
\putVtrianglep<\arrowtypea`\arrowtypeb`
\arrowtypec;\height>
(0,0)[#2`#3`#4;#5`#6`{#7}]
\enddiagram
}}

\def\putVtrianglepairp<#1>(#2)[#3;#4`#5`#6`#7`#8]{{
\settripairparms[#1]%
\setpos(#2)%
\settokens`#3`%
\advance\ypos by\height
\putmorphism(\xpos,\ypos)(1,-1)[`\tokend`{#6}]{\height}{\arrowtypec}l%
\puthmorphism(\xpos,\ypos)[\tokena`\tokenb`{#4}]{\height}{\arrowtypea}a%
\advance\xpos by\height
\puthmorphism(\xpos,\ypos)[\phantom{\tokenb}`\tokenc`{#5}]%
{\height}{\arrowtypeb}a%
\putvmorphism(\xpos,\ypos)[``{#7}]{\height}{\arrowtyped}m%
\advance\xpos by\height
\putmorphism(\xpos,\ypos)(-1,-1)[``{#8}]{\height}{\arrowtypee}r%
}}

\def\putVtrianglepair{\@ifnextchar <{\putVtrianglepairp}{\putVtrianglepairp%
    <\arrowtypea`\arrowtypeb`\arrowtypec`\arrowtyped`\arrowtypee;\height>}}
\def\Vtrianglepair{\@ifnextchar <{\Vtrianglepairp}{\Vtrianglepairp%
    <\arrowtypea`\arrowtypeb`\arrowtypec`\arrowtyped`\arrowtypee;\height>}}
\def\Vtrianglepairp<#1>[#2;#3`#4`#5`#6`#7]{{
\settripairparms[#1]
\settokens`#2`
\diagram
\putVtrianglepairp                             
<\arrowtypea`\arrowtypeb`\arrowtypec`
\arrowtyped`\arrowtypee;\height>
(0,0)[{#2};#3`#4`#5`#6`{#7}]
\enddiagram
}}

\def\putCtrianglep<#1>(#2,#3)[#4`#5`#6;#7`#8`#9]{{%
\settriparms[#1]%
\xpos=#2 \ypos=#3 \advance\ypos by\height
\putmorphism(\xpos,\ypos)(1,-1)[``{#9}]{\height}{\arrowtypec}l%
\advance\xpos by\height \advance\ypos by\height
\putmorphism(\xpos,\ypos)(-1,-1)[#4`#5`{#7}]{\height}{\arrowtypea}l%
{\multiply\height by 2
\putvmorphism(\xpos,\ypos)[`#6`{#8}]{\height}{\arrowtypeb}r}%
}}

\def\putCtriangle{\@ifnextchar <{\putCtrianglep}{\putCtrianglep
    <\arrowtypea`\arrowtypeb`\arrowtypec;\height>}}
\def\Ctriangle{\@ifnextchar <{\Ctrianglep}{\Ctrianglep
    <\arrowtypea`\arrowtypeb`\arrowtypec;\height>}}
\def\Ctrianglep<#1>[#2`#3`#4;#5`#6`#7]{{
\settriparms[#1]
\width=\height                               
\diagram
\putCtrianglep<\arrowtypea`\arrowtypeb`
\arrowtypec;\height>
(0,0)[#2`#3`#4;#5`#6`{#7}]
\enddiagram
}}                                           
\def\putDtrianglep<#1>(#2,#3)[#4`#5`#6;#7`#8`#9]{{%
\settriparms[#1]%
\xpos=#2 \ypos=#3 \advance\xpos by\height \advance\ypos by\height
\putmorphism(\xpos,\ypos)(-1,-1)[``{#9}]{\height}{\arrowtypec}r%
\advance\xpos by-\height \advance\ypos by\height
\putmorphism(\xpos,\ypos)(1,-1)[`#5`{#8}]{\height}{\arrowtypeb}r%
{\multiply\height by 2
\putvmorphism(\xpos,\ypos)[#4`#6`{#7}]{\height}{\arrowtypea}l}%
}}

\def\putDtriangle{\@ifnextchar <{\putDtrianglep}{\putDtrianglep
    <\arrowtypea`\arrowtypeb`\arrowtypec;\height>}}
\def\Dtriangle{\@ifnextchar <{\Dtrianglep}{\Dtrianglep
   <\arrowtypea`\arrowtypeb`\arrowtypec;\height>}}
\def\Dtrianglep<#1>[#2`#3`#4;#5`#6`#7]{{
\settriparms[#1]
\width=\height                              
\diagram
\putDtrianglep<\arrowtypea`\arrowtypeb`
\arrowtypec;\height>
(0,0)[#2`#3`#4;#5`#6`{#7}]
\enddiagram
}}                                          
\def\setrecparms[#1`#2]{\width=#1 \height=#2}%

\def\recursep<#1`#2>[#3;#4`#5`#6`#7`#8]{{\m@th
\width=#1 \height=#2 \settokens`#3`
\settowidth{\tempdimen}{$\tokena$} \ifdim\tempdimen=0pt
  \savebox{\tempboxa}{\hbox{$\tokenb$}}%
  \savebox{\tempboxb}{\hbox{$\tokend$}}%
  \savebox{\tempboxc}{\hbox{$#6$}}%
\else
  \savebox{\tempboxa}{\hbox{$\hbox{$\tokena$}\times\hbox{$\tokenb$}$}}%
  \savebox{\tempboxb}{\hbox{$\hbox{$\tokena$}\times\hbox{$\tokend$}$}}%
  \savebox{\tempboxc}{\hbox{$\hbox{$\tokena$}\times\hbox{$#6$}$}}%
\fi \ypos=\height \divide\ypos by 2 \xpos=\ypos \advance\xpos by
\width \bfig
\putCtrianglep<-1`1`1;\ypos>(0,0)[`\tokenc`;#5`#6`{#7}]%
\puthmorphism(\ypos,0)[\tokend`\usebox{\tempboxb}`{#8}]{\width}{-1}b%
\puthmorphism(\ypos,\height)[\tokenb`\usebox{\tempboxa}`{#4}]{\width}{-1}a%
\advance\ypos by \width
\putvmorphism(\ypos,\height)[``\usebox{\tempboxc}]{\height}1r%
\efig }}

\def\recurse{\@ifnextchar <{\recursep}{\recursep<\width`\height>}}

\def\puttwohmorphisms(#1,#2)[#3`#4;#5`#6]#7#8#9{{%
\puthmorphism(#1,#2)[#3`#4`]{#7}0a \ypos=#2 \advance\ypos by 20
\puthmorphism(#1,\ypos)[\phantom{#3}`\phantom{#4}`#5]{#7}{#8}a
\advance\ypos by -40
\puthmorphism(#1,\ypos)[\phantom{#3}`\phantom{#4}`#6]{#7}{#9}b }}

\def\puttwovmorphisms(#1,#2)[#3`#4;#5`#6]#7#8#9{{%
\putvmorphism(#1,#2)[#3`#4`]{#7}0a \xpos=#1 \advance\xpos by -20
\putvmorphism(\xpos,#2)[\phantom{#3}`\phantom{#4}`#5]{#7}{#8}l
\advance\xpos by 40
\putvmorphism(\xpos,#2)[\phantom{#3}`\phantom{#4}`#6]{#7}{#9}r }}

\def\puthcoequalizer(#1)[#2`#3`#4;#5`#6`#7]#8#9{{%
\setpos(#1)%
\puttwohmorphisms(\xpos,\ypos)[#2`#3;#5`#6]{#8}11%
\advance\xpos by #8
\puthmorphism(\xpos,\ypos)[\phantom{#3}`#4`#7]{#8}1{#9} }}

\def\putvcoequalizer(#1)[#2`#3`#4;#5`#6`#7]#8#9{{%
\setpos(#1)%
\puttwovmorphisms(\xpos,\ypos)[#2`#3;#5`#6]{#8}11%
\advance\ypos by -#8
\putvmorphism(\xpos,\ypos)[\phantom{#3}`#4`#7]{#8}1{#9} }}

\def\putthreehmorphisms(#1)[#2`#3;#4`#5`#6]#7(#8)#9{{%
\setpos(#1) \settypes(#8)
\if a#9 %
     \vertsize{\tempcounta}{#5}%
     \vertsize{\tempcountb}{#6}%
     \ifnum \tempcounta<\tempcountb \tempcounta=\tempcountb \fi
\else
     \vertsize{\tempcounta}{#4}%
     \vertsize{\tempcountb}{#5}%
     \ifnum \tempcounta<\tempcountb \tempcounta=\tempcountb \fi
\fi \advance \tempcounta by 60
\puthmorphism(\xpos,\ypos)[#2`#3`#5]{#7}{\arrowtypeb}{#9}
\advance\ypos by \tempcounta
\puthmorphism(\xpos,\ypos)[\phantom{#2}`\phantom{#3}`#4]{#7}{\arrowtypea}{#9}
\advance\ypos by -\tempcounta \advance\ypos by -\tempcounta
\puthmorphism(\xpos,\ypos)[\phantom{#2}`\phantom{#3}`#6]{#7}{\arrowtypec}{#9}
}}

\def\setarrowtoks[#1`#2`#3`#4`#5`#6]{%
\def\toka{#1}
\def\tokb{#2}
\def\tokc{#3}
\def\tokd{#4}
\def\toke{#5}
\def\tokf{#6}
}
\def\hex{\@ifnextchar <{\hexp}{\hexp<1000`400>}}
\def\hexp<#1`#2>[#3`#4`#5`#6`#7`#8;#9]{%
\setarrowtoks[#9] \yext=#2 \advance \yext by #2 \xext=#1
\advance\xext by \yext \bfig
\putCtriangle<-1`0`1;#2>(0,0)[`#5`;\tokb``\tokd] \xext=#1 \yext=#2
\advance \yext by #2
\putsquare<1`0`0`1;\xext`\yext>(#2,0)[#3`#4`#7`#8;\toka```\tokf]
\advance \xext by #2
\putDtriangle<0`1`-1;#2>(\xext,0)[`#6`;`\tokc`\toke] \efig }
\makeatother

\title{\bf\Large A Dodecalogue of Basic Didactics from Applications of\\ Abstract Differential Geometry to Quantum Gravity}

\author{Ioannis Raptis\thanks{EU Marie Curie Reintegration Research Fellow, Algebra and Geometry Section,
Department of Mathematics, University of Athens,
Panepistimioupolis, Athens 157 84, Greece; {\em and} Visiting
Researcher, Theoretical Physics Group, Blackett Laboratory,
Imperial College of Science, Technology and Medicine, Prince
Consort Road, South Kensington, London SW7 2BZ, UK; e-mail:
i.raptis@ic.ac.uk}}

\date{\today}

\begin{document}

{\catcode`\ =13\global\let =\ \catcode`\^^M=13
\gdef^^M{\par\noindent}}
\def\verbatim{\tt
\catcode`\^^M=13 \catcode`\ =13 \catcode`\\=12 \catcode`\{=12
\catcode`\}=12 \catcode`\_=12 \catcode`\^=12 \catcode`\&=12
\catcode`\~=12 \catcode`\#=12 \catcode`\%=12 \catcode`\$=12
\catcode`|=0 }

\maketitle

\pagestyle{myheadings}\markboth{\centerline {\small {\sc {Ioannis
Raptis}}}}{\centerline {\footnotesize {\sc {A Dodecalogue of ADG
Didactics on QG}}}}

\pagenumbering{arabic}

\begin{abstract}

\noindent {\small We summarize the twelve most important in our
view novel concepts that have arisen, based on results that have
been obtained, from various applications of Abstract Differential
Geometry (ADG) to Quantum Gravity (QG). The present document may
be used as a concise, yet informal, discursive and peripatetic
conceptual guide-{\it cum}-terminological glossary to the
voluminous technical research literature on the subject. In a
bonus section at the end, we dwell on the significance of
introducing new conceptual terminology in future QG research by
means of `poetic language'.}

\vskip 0.1in



\end{abstract}

\setlength{\textwidth}{15.0cm} 
\setlength{\oddsidemargin}{0cm}  
\setlength{\evensidemargin}{0cm} 
\setlength{\topskip}{0pt}  
\setlength{\textheight}{21.0cm} 
\setlength{\footskip}{-2cm}

\setlength{\topmargin}{0pt}

\section{A Dodecalogue of ADG-Didactics on QG}

Anastasios Mallios' Abstract Differential Geometry (ADG) has been
with us, at least in research monograph form, for almost a decade
now \cite{mall1}. During this time, it has enjoyed numerous
applications to quantum gauge theories and gravity
\cite{mall2,malros1,mall3,malros2,malros3,mall5,mall6,mall9,mall11,mall10,mall7,malrap1,malrap2,malrap3,malrap4,rap5,rap7,rap11,rap13},\footnote{In
the present text little reference to the `satellite' ({\it ie},
non-ADG based) QG literature, such as Loop Quantum Gravity (LQG)
or Causal Set (:Causet) Theory, is given. The reader should refer
to these papers for extensive citations of research monographs and
papers on QG. In \cite{malrap4,rap11} especially, the reader can
find fairly comprehensive and updated reference lists.}
culminating recently in another two-volume up to date treatise
\cite{mall4}.

Below, an informal {\it r\'esum\'e} of twelve novel concepts and
main results that have been gathered from applying ADG to QG is
given, a document that can at least provide the reader with a
handy guide, as well as a sketchy glossary of pivotal new terms,
to the existing literature `zoo' on the subject. The list of a
dozen items that follows is a distillation and summary of little,
albeit important, lessons that this author has learned from
employing, predominantly in collaboration with Mallios,
ADG-theoretic concepts, techniques and results to QG and quantum
gauge theories mainly in a finitistic-algebraic and categorical
setting
\cite{rap2,malrap1,malrap2,malrap3,malrap4,rap5,rap7,rap11,rap13}.\footnote{For
another fairly recent review on the subject, see \cite{rap11}. The
`{\em Glossary for ADG-Gravity}' appended at the end of
\cite{malrap4} may also be of help, although some of the terms and
definitions found there are presently ({\it ie}, two years later!)
slightly more refined, expanded and supported by further work that
has been done in the meantime.} Some of the twelve `lessonets'
(:`$\mu\alpha\theta\eta\mu\alpha\tau\iota\kappa\acute{\alpha}$'=Greek
for `{\em mathematics}'!)\footnote{In Greek, the word
`$\mu\acute{\alpha}\theta\eta\mu\alpha$' means `{\em lesson}';
hence, `mathematics' is used above as the diminutive of `{\em
lessons}' (:`$\mu\alpha\theta\acute{\eta}\mu\alpha\tau\alpha$').}
listed below are closely related to each other.

\begin{enumerate}

\item {\bf Background spacetime manifoldlessness.}
ADG-gravity\footnote{Following
\cite{malrap3,malrap4,rap5,rap7,rap11,rap13}, we subsume the
ADG-theoretic formulation of gravity (classical and/or quantum,
with no significant distinction between these two epithets---see
item 10 on the list)
\cite{mall3,malros1,malros2,malros3,malrap1,malrap2,malrap3,mall4,rap5,rap7,rap11,mall4,mall9,mall11,mall7}
under the term `{\em ADG-gravity}'.} is manifestly background
spacetime manifold independent. There is no smooth locally
Euclidean base space  involved in formulating the (vacuum)
Einstein equations, hence no space(time) interpretation of that
base either.

\item {\bf Half-Order Formalism.} The sole dynamical variable in
ADG-gravity is an algebraic $\struc$-connection field
$\conn$\footnote{Where $\struc$ is  a sheaf of unital, commutative
and associative $\mathbb{K}$-algebras ($\mathbb{K}=\R ,\com$)
called the {\em structure sheaf} of generalized arithmetics or
coordinates. $\struc$ is chosen by the theorist (or the
`experimenter'/`observer'!) and it generalizes the usual structure
sheaf $\smooth_{M}$ of germs of smooth (usually taken to be
$\R$-valued) coordinate functions on a differential manifold $M$.}
acting (on the local sections of) a vector sheaf $\modl$ defined
on an in principle arbitrary topological space $X$. The physical
kinematical configuration space in the theory is the moduli space
$\sconn_{\struc}(\modl)/\aut\modl$ of the affine space of
connections $\sconn$ modulo the (local) gauge transformations in
the principal sheaf $\aut\modl$. The ADG-formalism on gravity is
called half-order formalism in order to distinguish it from the
first-order one of Palatini-Ashtekar, and from the original
second-order one of Einstein, both of which depend on a background
smooth manifold for their differential geometric expression. From
the ADG-theoretic vantage, gravity is regarded as a pure gauge
theory since only the connection, and not the smooth metric (or
equivalently, the smooth tetrad field), is a dynamical variable.
It follows that the connection $\conn$ represents the
gravito-inertial field and, unlike the $g_{\mu\nu}$ of GR, {\em
not} the chrono-geometrical structure. There is no `spacetime
geometry' in ADG-gravity, or rather more mildly, if there is any
`space' (:`geometry') at all, it is already encoded in the
$\struc$ chosen.\footnote{In the usual case
$\struc\equiv\smooth_{M}$, we have in mind here the notion of
Gel'fand duality whereby the differential manifold $M$ is the
spectrum of the (non-normed) topological algebra sheaf
$\smooth_{X}$.} `Geometry' (or indeed, `spacetime') is completely
encoded in our (generalized) measurements in $\struc$. There is no
geometry without measurement, without the production (:recording)
of numbers of some sort.\footnote{Normally, we assume that we
measure real ($\R$) numbers.} At the same time, (the products of)
measurements (:numbers) are our own actions (and numbers our own
artifacts/inventions), hence no physical reality, and no
interpretation as the gravitational field living `out there',
should be given to the spacetime metric, like in the original
formulation of GR. This is consistent with our viewing gravity as
a pure gauge theory---{\it ie}, that the gravitational field is
simply the connection $\conn$.

\item {\bf All differential geometry boils down to $\struc$.} To
be sure, there would be no {\em differential} geometry without a
differential, in fact, without $\conn$, and $\conn$ has $\struc$
as its domain of definition. There is an apparent oxymoron here:
we are inclined to give physical reality (:independence from our
observations/measurements in $\struc$) to the gravitational
connection field $\conn$, yet here we seem to claim that $\struc$
is vital for its very definition. There is no paradox: the
ADG-vacuum Einstein gravitational equations are expressions
involving the {\em curvature} of $\conn$, which is a `{\em
geometrical object} (:an
$\otimes_{\struc}$-tensor,\footnote{$\otimes_{\struc}$ is the
homological tensor product functor between the (vector) sheaf
categories involved in ADG-gravity.} or equivalently, an
$\struc$-sheaf morphism) in the theory, hence the dynamics is
functorial relative to $\struc$. On the other hand, $\conn$ is
{\em not} an $\otimes_{\struc}$-tensor, hence it eludes our
measurements in $\struc$. $\struc$ is simply employed to
effectuate (:geometrically represent) $\conn$ on a vector sheaf
$\modl$,\footnote{By definition, $\modl$ is locally of the form
$\struc^{n}$ ($n$ a positive integer called the rank of the vector
sheaf $\modl$).} which is the carrier (:representation-action)
space of $\conn$.

\item {\bf Gauge Theory of the Third Kind.} The half-order
ADG-gravity, although by definition (and sheaf-theoretic
construction!) a local gauge theory, it is not local in the usual
sense of the current so-called gauge theories of the second kind
since no background spacetime manifold is employed to localize the
ADG-gravitational field and the symmetries of the dynamical
equations that it obeys (:the vacuum Einstein equations). It
follows that the group $\mathrm{Diff}(M)$ of smooth external
spacetime (as opposed to internal gauge) symmetries does not
figure in the theory, and all the `symmetries'  of the vacuum
gravitational dynamics in $\aut\modl$ are `internal' to the
ADG-gravitational field $(\modl ,\conn)$.\footnote{Of course, in
case one chooses $\struc\equiv\smooth_{X}$ ({\it ie}, $X\equiv M$
by Gel'fand duality), $\mathrm{Aut}(M):=\mathrm{Diff}(M)$ and one
falls back to the usual scenario (:the manifold based GR).}

\item {\bf Unconstrained Gauge Gravity.} From the above, it
follows that ADG-gravity is an external smooth spacetime
unconstrained pure gauge theory. The primary, externally imposed
(by the theorist, or the `experimenter')  spacetime gauge
constraints have been eliminated in ADG-gravity, so that what
remains is the dynamically autonomous ADG-gravitational connection
field $\conn$.

\item {\bf Algebraicity, Representation, Categoricity and
Functoriality.} ADG-gravity is a purely algebraic
(:sheaf-theoretic) scheme. Its main structural element, the
algebraic $\struc$-connection field $\conn$ is a linear Leibnizian
sheaf morphism acting categorically on $\modl$'s local sections to
change them dynamically. In turn, $\modl$ is regarded as the
associated (:representation or carrier/action) sheaf of the
principal group sheaf $\aut\modl$, while from a geometric
(pre)quantization and second quantization vantage, its local
section represent local quantum-particle states of the ADG-fields
involved. The ADG-field---by definition, the pair $(\modl
,\conn)$---has thus been coined `particle-field pair', and the
vacuum Einstein equations $\ricci(\conn)[\modl]=0$ have a
straightforward geometrical interpretation: the connection acts
via its (Ricci) curvature on the local particle states (:local
sections) of $\modl$ to change them dynamically. Moreover, since
the said curvature is an $\struc$-morphism (or equivalently, an
$\otimes_{\struc}$-tensor, with $\otimes_{\struc}$ the homological
tensor product functor), the dynamics is functorial relative to
our generalized coordinate  measurements in (:coordinatizations of
the connection field $\conn$ by) $\struc$.\footnote{In our scheme,
coordinatizing the algebraic connection field $\conn$ is
tantamount to representing it on $\modl$, which by definition is
locally of the form $\struc^{n}$ (:a locally free $\struc$-module
of finite rank $n$). In turn, $\modl$ is the associated
(:representation) sheaf of $\aut\modl$.}  This
$\struc$-functoriality (or equivalently, the
$\aut\modl$-invariance) of the vacuum Einstein dynamics is the
ADG-analogue of the Principle of General Covariance (PGC) of the
$\smooth$-smooth manifold based  GR in which the PGC is
effectuated via $\mathrm{Diff}(M)$.\footnote{To be sure, since  by
definition $\modl$ is locally a finite power of $\struc$, $\modl
|_{U}:=\struc^{n}(U)$, $\aut\modl$ is also locally isomorphic to
the general coordinate group sheaf:
$\aut\modl(U)=\mathcal{G}\mathcal{L}(n,\struc(U))\equiv
M_{n}(\struc(U))^{\bull},~(U\subset X$ open).} All in all, ADG is
an entirely algebraic way of doing differential geometry whereby
all that is of technical import and physical significance is the
algebraic relations between the sections of the sheaves involved,
without the mediation in the guise of (smooth) coordinates of a
background geometrical differential (spacetime) manifold.

\item {\bf Third Quantization.} Third quantization is ADG-field
quantization. As noted above, the ADG-fields are defined to be
pairs $\field :=(\modl ,\conn)$, and 3rd-quantization pertains to
setting up non-trivial local commutation relations between certain
characteristic local (:differential) forms that uniquely
characterize sheaf cohomologically the vector sheaves $\modl$ and
the connections $\conn$ on them. Based on the ADG-field semantics
coming from applications of ADG to geometric (pre)quantization and
2nd-quantization, the said characteristic forms may be interpreted
as abstract particle-position and field-momentum `determinations'
(:operations or actions), so that their commutation relations can
be regarded as abstract canonical-type of local Heisenberg
uncertainty relations. Moreover, since the ADG-formalism is
inherently background spacetime manifoldless, 3rd-quantization of
ADG-gravity, in contradistinction to the connection, but also
manifold, based Loop Quantization in canonical QGR, it does not
involve at all quantization of the spacetime continuum itself.
Rather, it is an autonomous quantization scenario for the
ADG-fields $\field$ `in-themselves', as it involves solely their
abstract particle-position ($\modl$) and field-momentum ($\conn$)
parts. 3rd-quantization is manifestly $\struc$-functorial, hence
expressly fully (:generally) covariant, whereas it takes a lot of
effort to secure that the usual canonical
$\mathrm{Diff}(M)$-constrained Dirac quantization of GR be
(generally) covariant (even in the Ashtekar variables formulation
of GR, which significantly simplifies the primary
spacetime-diffeomorphism constraints).  Finally, `observables' in
3rd-quantized ADG-vacuum Einstein gravity are all the
$\struc$-morphisms ({\it alias}, $\otimes_{\struc}$-functors like
the curvature of $\conn$)---the geometrical (:$\struc$-measurable)
dynamical objects.

\item {\bf No Singularities and No Infinities'.} It is plain that
in a purely algebraico-categorical scheme such as ADG-gravity,
there are no infinities, since the infinitary (:limiting)
processes of the usual continuum mediated Analysis that we have
hitherto used in QG research are replaced by Algebra (:the
dynamical relations between the geometrical objects in the
theory---the $\otimes_{\struc}$-tensors or
$\struc$-morphisms),\footnote{More generally, in ADG we have
effectively replaced, in a {\em differential} geometric context,
Analysis by sheaf cohomology ({\it ie}, `space' by homological
algebra) in a way analogous to what Grothendieck did in {\em
algebraic} geometry.} and there are no (analytic) infinities in
the latter. Furthermore, singularities of all sorts, even
non-linear distributional ones, are seen to be absorbed (or
integrated) in $\struc$ so that by virtue of the
$\struc$-functoriality of the ADG-gravitational dynamics, the
latter is seen {\em not} to break down in any differential
geometric sense in the vicinity of the singular {\em loci}. The
ADG-gravitational field $(\modl ,\conn)$ (and the
$\struc$-functorial vacuum Einstein equations that it defines via
its curvature $\struc$-morphism)  `sees' through the singularities
which are embodied in $\struc$.

\item {\bf No Fundamental Scales, No Continuum/Discretum
Dichotomy.} Much in the same way that we cannot accept that there
are singularities and infinities in Nature, or that there is a
spacetime (continuum) separate from ({\it ie}, externally
prescribed to) the dynamical fields themselves---a continuum that
is responsible for the singularities and their associated
(analytic) infinities in the first place---we cannot accept that
there are fundamental spacetime scales above which the law of GR
holds, but below which another set of laws (those of the QG we are
after) does. Such regularization scales are normally posited in
order to make sense out of the non-renormalizable  infinities of
our manifold based Analysis and have nothing to do with Nature
({\it ie}, with the dynamical field laws themselves).

{\it Mutatis mutandis} for the continuum/discretum dichotomy in
ADG-gravity: if one `concocts' supposedly fundamental spacetime
scales, such as Planck's, out of the universal constants ({\it
eg}, $\hbar ,c, G$) which in the theory are represented by
(global) sections of the constant sheaf $\mathbf{K}$, the
$\struc$- and, plainly, the $\mathbf{K}$-functoriality\footnote{It
is tacitly assumed that
$\mathbf{K}\stackrel{\subset}{\rightarrow}\struc$.} of the
ADG-gravitational dynamics indicates that the
theory\footnote{Implicit here is the `{\it credo}' that a physical
theory is nothing else but the dynamical principles and structures
that it assumes, as well as the dynamical laws that it is able to
formulate according to them.} recognizes no such cut-off scales
delimiting and separating (spacetime) continuum from discretum
based scenaria.\footnote{For example, the continuum based GR
distinguished from the `discrete' causal set (causet) theory.}
Since no base spacetime is involved in ADG-gravity, it is
meaningless to make such continuum/discretum distinctions either.
It follows that top-down/bottom-up distinctions between various
approaches to QG, such as LQG and Causet Theory respectively, lose
their meaning in ADG-gravity.\footnote{On the other hand, by
ADG-means we are able pass through the horns of the
continuum/discretum dilemma, bring together and fuse to a certain
extent various central structures involved in the continuum based
LQG and the discrete causet theory
\cite{rap1,malrap1,malrap2,malrap3,malrap4,rap5,rap7}.} The
field-law of gravity appears to be more unified (:`unitary') in
ADG-gravity than in other approaches to QG that {\it a priori}
suppose a (background spacetime) continuum/discretum distinction.
To stress it again, ADG-gravity has no background continuum or
discretum spacetime commitments.

\item {\bf No Classical/Quantum Schism.} Perhaps the most radical
`no' of the ADG-gravitational didactics is that renouncing the
classical/quantum separation (:Heisenberg's {\it schnitt}).
Generally speaking, it seems intuitively naive and contrived to
assume that there are classical physical systems and theories
describing them, and moreover, that a formal procedure of
quantization should take us to their fundamental quantum
correspondents. Conversely, it seems artificial and presumptuous
to demand that any quantum description of a physical situation
should yield at an appropriate so-called correspondence limit a
classical description (of the same system!), let alone to adopt as
a criterion for the acceptability of a quantum theory whether it
yields, in a suitable `limit' sense, its classical counterpart as
a coarse, effective theory. There is a psychological factor
inherent in our `correspondence principle' conservatism: we seem
to trust more `classical' than `quantum' theories, because the
former have been more experimentally vindicated (or at least they
have been around longer than quantum ones, and they have been
longer worked out mathematically and better understood
conceptually).  Accordingly, we demand from our quantum theories
to set, via a suitable correspondence principle, one foot on the
classical domain for stability and acceptability's sake. By
contrast, in the past we have argued extensively that our
ADG-gravity is already (3rd) quantum since the ADG-gravitational
field $(\modl ,\conn)$, and the law that it defines, has {\it ab
initio} quantum traits inherent in its structure. Thus, the
ADG-gravitational field is in no need of of a (formal) process of
quantization, while 3rd-quantization is a procedure of explicating
its `self-quantum' nature (not a quantization of some `classical'
sort of ADG-gravity, which does not exist from our
perspective).\footnote{To be sure, as a formal `correspondence
principle' in ADG-gravity one could take the assumption
$\struc\equiv\smooth_{M}$, but this should not be regarded as a
transition from quantum to classical ADG-gravity: in 3rd-quantum
ADG-gravity all structure sheaves are on a par.}

\item {\bf ADG-Field Monadology, Unitarity, Quantum Dynamical
Autonomy and Synvariance.} From the ADG-perspective, the field
$\field :=(\modl ,\conn)$ is, to paraphrase Einstein
\cite{einst3}, ``{\em an independent, not further reducible
fundamental concept}''. In fact, it is an `entelechic',
dynamically autonomous quantum structure, since on the one hand
the dynamical equations (Einstein's equations) are defined solely
in terms of it, while on the other, the 3rd-quantum Heisenberg
uncertainty relations mentioned above are defined
intrinsically---{\it ie}, only in terms of its two constituent
parts: $\modl$ and $\conn$.\footnote{For this, the field $\field
:=(\modl ,\conn)$ has been coined `self-quantum'.} As mentioned
before, since no spacetime (continuous or discrete) external to
the ADG-gravitational field is posited, the
$\mathrm{Diff}(M)$-modelled PGC of GR is hereby replaced by
$\aut_{\struc}\modl$ so that `covariance' is replaced by
`synvariance': the ADG-gravitational field varies dynamically
`in-itself', not relative to an external (`parameter')
space(time). In this respect it reminds one of the Leibnizian
monads \cite{leibniz}, which are entelechic, windowless entities.;
although, the ADG-fields are relational (:algebraic) entities that
can interact with each other. Moreover, this external
spacetimelessness makes the ADG-gravitational field an
unconstrained, pure gauge quantum system (gauge field theory of
the 3rd-kind; 3rd-quantization).

Finally, in the context of ADG, the epithet `unitary' to `field
theory' pertains less to Einstein's original vision of `one single
field encompassing all forces'\footnote{Indeed, currently
`unitary' has degenerated to `unified' in the sense above: one
field for all forces.} and more to (also Einstein's, less popular
however, vision of) a `total field' obeying (defining) total
dynamical (differential) equations that are `free' from ({\it ie},
in no way impeded by, let alone breaking down in the presence of)
singularities, it incorporates its particle-quanta as
singularities in the law (differential equation) that it defines,
and (as a bonus to Einstein's vision---and something that Einstein
could not have possibly envisioned in the manifold and {\it in
extenso} CDG-based field theory that he was espousing) the
background spacetime continuum plays no role whatsoever in the
field's autonomous dynamics (`autodynamics'). {\it In summa}, a
genuinely unitary field theory in our ADG-theoretic sense is
concerned only with the field, the whole field and nothing but the
field.

\item {\bf The Principles of Relativity of Differentiability,
ADG-Field Realism and Field-Solipsism.} Since the basic tenet of
ADG is that all differential geometry boils down to (our choice
of) $\struc$, the Principle of Algebraic Relativity of
Differentiability (PARD) pertains to different choices (on the
part of the, external to the ADG-gravitational $\struc$-connection
field $\conn$, theorist/observer/experimenter) of structure sheaf
$\struc$ of generalized arithmetics (:coordinates or
measurements), hence of different algebras of differentiable
functions and different representation vector sheaves carrying the
connection field $\conn$, while at the same time leaving the
dynamical equations form-invariant.\footnote{For example, such
changes (in our choice) of $\struc$ may be made, because we might
wish to include an occurring singularity in a new functional
structure sheaf that the old one did not embody.}  Categorically,
PARD is modelled after natural transformation-type of changes,
which leave `invariant' the $\struc$-functorial vacuum
ADG-gravitational dynamics.

A corollary of PARD is the Principle of ADG-Field Realism (PFR):
one is in principle  free to choose (according to the physical
situation/problem one encounters) different $\struc$s to `measure'
and geometrically represent (on
$\modl:\stackrel{\mathrm{loc.}}{=}\struc^{n}$) , but the field and
the law that it obeys (better, defines) remain unaffected by such
changes. The field exists `out there', independently of our
`measurements' (of it) in $\struc$ and, accordingly, of our
geometrical representations in $\modl$.

This ADG-field realism, together with the aforesaid quantum
dynamical ADG-field autonomy, make us coin the philosophy
underlying ADG-field theory `{\em ADG-field solipsism}. For,
solipsism is `pure realism'.

\end{enumerate}

\section{Poetry in Motion and in Action: the Future of QG Research}

\paragraph{Descending to the quantum deep: the `experience-to-theoretical physics-to-mathematics-to-philosophy-to-poetry' ascension.}
In QG research, because of the glaring absence of experimental
data (in fact, of any controlled laboratory
experiments!)\footnote{Although we are passive receptors of
cosmological data from the early universe.} to verify (or more
importantly, to falsify!) our theories, the
theoretical/mathematical physicist finds herself in the fortuitous
position of being free to roam in unconstrained, uninhibited
theory making, with sole guiding tools `aesthetic' elements such
as conceptual simplicity, economy and beauty, backed by
mathematical abstraction, generality, rigor and logical
consistency. This has been appreciated as early as Dirac's days
\cite{dirac3}, who implored theoretical physicists to explore and
use all the {\em mathematical} resources at their disposal, and
divesting for a while experiments of their theory checking and
guiding  role.

Faddeev, for example, maintained recently \cite{faddeev} that we
should finally break away from the classical theory-developmental
route followed so far by theoretical physics, which can be
schematically represented by the cycle:

{\small
$$
\mathrm{experiments}~\mapto~\mathrm{predictions}\mapto\mathrm{mathematical~
elaborations/formulations}\mapto\mathrm{further~experiments}
$$
}

\noindent and instead implore all our mathematical resources to
plough deeper into the foundations of `physical reality', leaving
experiments (and experimentalists!) to `catch up' with the new
mathematics (and with theoreticians!), not the other way round. In
this regard, we would like to borrow from \cite{faddeev} some
telling remarks made by Dirac from the aforementioned paper
\cite{dirac3}:\footnote{The quotation below is split into two
paragraphs (I and II), on which we comment separately after it.}

\begin{quotation}
\noindent ``{\small ...The steady progress of physics requires for
its theoretical foundation a mathematics that gets continually
more advanced. This is only natural and to be expected. What,
however, was not expected by the scientific workers of the last
century was the particular form that the line of advancement of
the mathematics would take, namely, it was expected that the
mathematics would get more complicated, but would rest on a
permanent basis of axioms and definitions, {\em while actually the
modern physical developments have required a mathematics that
continually shifts its foundation and gets more abstract...It
seems likely that this process of increasing abstraction will
continue in the future and that advance in physics is to be
associated with a continual modification and generalization of the
axioms at the base of mathematics rather than with logical
development of any one mathematical scheme on a fixed
foundation.}\footnote{Our emphasis.} {\bf (I)}

There are at present fundamental problems in theoretical physics
awaiting solution [...]\footnote{Dirac here mentions a couple of
outstanding mathematical physics problems of his times. We have
omitted them.}the solution of which problems will presumably
require a more drastic revision of our fundamental concepts than
any that have gone before. Quite likely these changes will be so
great that it will be beyond the power of human intelligence to
get the necessary new ideas by direct attempt to formulate the
experimental data in mathematical terms. The theoretical worker in
the future will therefore have to proceed in a more indirect way.
{\em The most powerful method of advance that can be suggested at
present is to employ all the resources of pure mathematics in
attempts to perfect and generalise the mathematical formalism that
forms the existing basis of theoretical physics, and {\sl
after}\footnote{Dirac's own emphasis.} each success in this
direction, to try to interpret the new mathematical features in
terms of physical entities}\footnote{Again, our emphasis
throughout.}...} {\bf (II)}''
\end{quotation}

\begin{itemize}

\item {\bf (I)} The words from this paragraph to be highlighted
with ADG-gravity in mind are: `{\em a mathematics that gets more
abstract}' and `{\em advance in physics is to be associated with a
continual process of abstraction {\rm [leading to a]} modification
and generalization of the axioms at the base of mathematics}'.
Indeed, the axiomatic ADG essentially involves an abstraction of
the fundamental notions of modern differential geometry ({\it eg},
connection), resulting in an entirely algebraic (:sheaf-theoretic)
modification and generalization of the latter's basic axioms
\cite{mall1,mall2,mall4}. And it is precisely this abstract and
generalized character of ADG that makes us hope that its
application could advance significantly (theoretical) physics, and
in particular, QG research. For, to quote again Einstein, in the
quantum deep we must look for ``{\em a purely algebraic method for
the description of reality}'' \cite{einst3}. \footnote{Alas, for
Einstein, the continuum spacetime and {\it in extenso} CDG-based
field theory was simply incompatible with the finitistic-algebraic
quantum theory \cite{stachel}, a divide that ADG has come a long
way to finally bridge
\cite{malrap1,malrap2,malrap3,malrap4,rap5,rap7,rap11,rap13}.}

\item {\bf (II)} In this paragraph, apart from breaking from the
traditional cycle `experiment-theory-more experiment' mentioned
above ({\it ie}, Dirac's anticipation that `{\em new ideas {\rm
[won't come]} by direct attempts to formulate the experimental
data in mathematical terms}'), what should be highlighted is on
the one hand Dirac's prompting us `{\em to generalize the
mathematical formalism that forms the existing basis of
theoretical physics}', and on the other, `{\em to try to interpret
the new mathematical features in terms of physical entities}'.
Again, ADG comes to fulfill Dirac's vision, since {\em the} (or at
least the bigger part of the) mathematics that lies at the heart
of current theoretical physics---namely, (the formalism of) {\em
differential geometry} ({\it ie}, CDG on smooth manifolds)---is
abstracted and generalized, while {\em after} this generalization
has been achieved, the physical application and interpretation (of
ADG's novel concepts and features) has been carried out,
especially in the theoretical physics' field of quantum gauge
theories and gravity research. We believe that this is `{\em a
powerful method of advance}' indeed.

\end{itemize}

\noindent However, this too is not enough. Existing mathematical
concepts, structures and techniques also come hand in hand with
implicit assumptions, hidden preconceptions and prejudices
associated with their historical development, {\it ie}, with past
problems other than QG(!) that they were invented in order to
formulate, tackle and (re)solve. Such preconceptions are very hard
to forget at the primary stages of theory making, let alone to
shed them altogether, especially when they have proved to be
experimentally successful in the past. Again Einstein, for
example, has given us a warning call regarding our almost
religious abiding by old, tried-and-tested concepts \cite{einst7}:

\begin{quotation}
\noindent ``{\small ...Concepts which have proved useful for
ordering things easily assume so great an authority over us, that
we forget their terrestrial origin and accept them as unalterable
facts. They then become labelled as `conceptual necessities', `a
priori situations', etc.\footnote{Think for instance of the
apparently fundamental notion of the `{\em spacetime continuum}':
``{\em time and space are modes by which we think, not conditions
in which we live}'' (as quoted by Manin in \cite{manin}). }The
road of scientific progress is frequently blocked for long periods
by such errors. It is therefore not just an idle game to exercise
our ability to analyze familiar concepts, and to demonstrate the
conditions on which their justification and usefulness depend, and
the way in which these developed, little by little...}"
\end{quotation}

For this, few people have suggested to go even a bit further, past
mathematics, and into the realm of {\em philosophy} to look for
novel QG research resources. 't Hooft, for example \cite{thooft},
insists that:

\begin{quotation}

\noindent ``{\small ...The problems of quantum gravity are much
more than purely technical ones. They touch upon very essential
philosophical issues...}''

\end{quotation}

\noindent For us, this will not suffice either. Philosophy too
comes burdened with a host of {\it a priori} concepts and
assumptions.\footnote{Especially the nowadays academic `Philosophy
of Science' \cite{sklar,clark}, which appears to be heavily
(almost paracytically!) dependent on the concepts, techniques,
results and current developments in science (and in particular, in
theoretical physics and applied mathematics).} Paraphrasing
Finkelstein \cite{drf}, ``{\em in the quantum deep one must travel
light}''. Alas, perhaps because of a deep psychological tendency
towards security (and an instinctive, biological one, for survival
\cite{wheat}), we tend to abide by what we already know and
(think) we understand (or believe to have a firm hold of backed by
numerous practical applications), and we take few `conservative
risks' (pun intended) towards standing bare, ignorant (but,
exactly thanks to this ignorance, uninhibited and unbiased!)
before Nature. This primordial fear of the unknown must be
overcome---at least it should be soothed by the Socratic stance
that, {\em anyway, the only thing that we know for sure is that we
know almost nothing}---and a way of achieving this is by engaging
into imaginative, creative {\em poetic activity} where there is
plenty of leeway for `trial-and-error' and a lot of room for
iconoclastic, unconventional and adventurous ideas that are
unburdened by ancestral theoretical demands or traditional
conventions..

Indeed, granted that QG pushes us back to theorizing about the
archegonal acts of the World, what better means other than poetry
(with its analogies, metaphors and allegories) do we possess for
exploring, conceptually afresh and without {\it a priori}
commitments---ultimately, to deconstruct and reconstruct anew
\cite{plotnitsky}---the strange,\footnote{`Strange', of course,
relative to what we already (think we) know!} uncharted QG
landscape? Kandinsky's words echo ecophantically here \cite{kand}:

\vskip 0.1in

\centerline{``{\em Poetry brings us closer to the Creator.}''}

\vskip 0.1in

Especially regarding the unfamiliar realm of the quantum, we read
from \cite{midgley} (reading from \cite{tolstoy}):

\begin{quotation}
\noindent ``{\small ...In the first forty years of the twentieth
century, our vision of the physical world changed radically and
irretrievably. Atoms could behave like solid matter or like waves,
they were made of particles with strange top-like properties, with
nuclei which could disintegrate spontaneously, and, perhaps, set
up chains of disintegration themselves. For many, the most
interesting implication of all this new knowledge was, and still
is, philosophical. We have understood that our intuitive ideas of
what is possible and what is not---our common sense---are a result
of the conditioning of our minds by sense-experiences. {\em We
have had to change our ideas of what understanding consists
in}.\footnote{Midgley's emphasis.} As Bohr said, `{\sl When it
comes to atoms, language can only be used as in poetry. The poet,
too, is not nearly so concerned with describing facts as with
creating images}.'\footnote{Our {\sl emphasis}.} The same is true
of cosmological models, curved spaces and exploding universe. {\em
Images and analogies are the keys}.\footnote{Midgley's emphasis,
and mine.} Not you, not I, not Einstein could interpret the
universe in terms wholly related to our senses. Not that it is
incomprehensible, no. {\sl But we must learn to ignore our
preconceptions concerning space, time and matter, abandon the use
of everyday language and resort to metaphor. \underline{We must
try to think like poets}...''}\footnote{{\sl Emphasis} (and
underlining) is all ours.}}
\end{quotation}

\noindent What we have in mind here is that, in order to see and
tackle the problem of QG afresh, we must foremost be able to sort
of `(re)create it from scratch', forgetting for a while the
voluminous body of work---the various theoretical `evidence' that
different approaches to QG provide us with---that has been
gathered over the last 70+ years of research on it. The spirit of
Feynman comes to mind:

\vskip 0.1in \centerline{``{\em What I cannot create, I do not
understand.}'' \cite{feyn}\footnote{In the `{\em Quantum Gravity}'
prologue by Brian Hatfield.}} \vskip 0.1in

Of course, by `poetry' above all we mean {\em creation of new
conceptual terminology within a novel theoretical and technical
framework'}. In this respect our ADG-inspired QG research
endeavors are akin (in spirit at least) to how Feynman wished to
tackle the problem of QG. As Hatfield accounts in the prologue to
\cite{feyn}:

\begin{quotation}
\noindent ``{\small ...Thus it is no surprise that Feynman would
recreate general relativity from a non-geometrical viewpoint. The
practical side of this approach is that one does not have to learn
some `fancy-schmanzy' (as he liked to call it) differential
geometry in order to study gravitational physics. (Instead, one
would just have to learn some quantum field theory.) However, when
the ultimate goal is to quantize gravity, Feynman felt that the
geometrical interpretation just stood in the way.\footnote{Recall
in this respect from \cite{feyn3}: ``{\em the simple ideas of
geometry, extended down to infinitely small, are wrong}''; and
more recently Isham \cite{isham}: ``{\em ...at the Planck-length
scale, differential geometry is simply incompatible with quantum
theory...[so that] one will not be able to use differential
geometry in the true quantum-gravity theory}''.} From the field
theoretic viewpoint, one could avoid actually defining---up
front---the physical meaning of quantum geometry, fluctuating
topology, space-time foam, {\it etc.}, and instead look for the
geometrical meaning after quantization...Feynman certainly felt
that the geometrical interpretation is marvellous, `{\em but the
fact that a massless spin-$2$ field can be interpreted as a metric
was simply a coincidence that might be understood as representing
some kind of gauge invariance'}\footnote{Our emphasis of Feynman's
words as quoted by Hatfield.}...}''
\end{quotation}

\noindent Similarly, within the novel theoretical framework of
ADG, ADG-gravity recreates GR purely gauge-theoretically; albeit,
unlike Feynman, not by doing away with differential geometry
altogether, but by (re)developing and casting the latter in an
entirely algebraic and background (spacetime manifold) independent
fashion, an achievement that suits perfectly current QG research
trends.

In this respect, it is perhaps more important to stress that ADG
is not so much a {\em new} theory of DG---the main `{\em
mathematical formalism that forms the existing basis of
theoretical physics}', following Dirac's expression earlier---but
a theoretical framework that abstracts, generalizes, revises and
recasts the existing CDG by isolating and capitalizing on its
fundamental, {\em essentially algebraic} (:`relational', in a
Leibnizian sense) features, which are not dependent at all on a
background locally Euclidean geometrical `space(time)'
(:manifold). In a way, from the novel viewpoint of ADG, we see
`old' and `stale' problems ({\it eg}, the $\smooth$-singularities
of the manifold and CDG based GR) with `new' and `fresh' eyes.
Schopenhauer's words from \cite{schopenhauer} immediately spring
to mind:

\begin{quotation}
\noindent ``{\small\em ...Thus, the task is not so much to see
what no one has yet seen, but to think what nobody yet has thought
about that which everybody sees\footnote{All emphasis is
ours.}...}''
\end{quotation}

\paragraph{On the `idiosyncratic' terminology side.} The novel perspective on gravity that ADG enables us to entertain
is inevitably accompanied by {\em new terminology}. We have thus
not refrained from engaging into vigorous poetic, `{\em
lexiplastic}' activity, so that the dodecalogue above abounds with
new, `idiosyncratic' terms for novel concepts hitherto not
encountered in the standard theoretical physics' jargon and
literature, such as `{\em gauge theory of the third kind}', `{\em
third quantization}', `{\em synvariance}' and `{\em
autodynamics}', to name a few.

In this respect, we align ourselves with Wallace Stevens' words in
\cite{stevens}:

\vskip 0.1in

\centerline{``{\small ...Progress in any aspect is a movement
through changes in terminology...}''}

\vskip 0.1in

\noindent with the `changes in terminology' in our case being not
just superficial (:formal) `nominal' ones introduced  as it were
for `flash, effect and decor', but necessary ones coming from {\em
a significant change in basic theoretical framework for viewing
and actually doing DG in QG}: from the usual geometrical manifold
based one (CDG), to the background manifoldless and purely
algebraic (:sheaf-theoretic) one of ADG.

\paragraph{The bottom line is a verse: a Word for the World.} `{\em In the beginning was the Word}',
thus the ultimate task for future QG (re)search is to find the
right `words' to begin our theory making about the very beginning
of the World. For, to quote Bohr (as quoted in \cite{au}):

 \begin{quotation}
\noindent ``...{\small It is wrong to think that the task of
physics is to point out how nature is. {\em Physics concerns what
we can say about nature}}...''\footnote{Our emphasis. What could
baffle the reader here is the following apparent oxymoron: while
on the one hand we seem to advocate the aforesaid principle of
ADG-field realism (maintaining that the connection field $\conn$
exists `out there' independently of us experimenters,
measurers/geometers and theoreticians), on the other we endorse
Bohr's dictum above. Again, there's no paradox here: what {\em we}
can say about Nature ({\it ie}, in this case, about the field
$\conn$) is all encoded in the generalized arithmetics $\struc$
that {\em we} choose to represent it (on $\modl$). However, the
$\struc$-functoriality of the dynamics secures the independence of
the (dynamics of the) field from our generalized measurements (and
hence from our geometrical representations, {\it eg}, `spacetime')
in $\struc$ (and {\it in extenso} $\modl$, which is locally a
power of $\struc$).}
\end{quotation}

\noindent As Finkelstein notes,\footnote{In an early draft of
\cite{drf} given to this author back in 1993.}

\begin{quotation}
\noindent ``{\small ...The fully quantum theory lies somewhere
within the theorizing activity of the human race itself, or the
subspecies of physicists, regarded as a quantum system. If this is
indeed a quantum entity, then the goal of knowing it completely is
a Cartesian fantasy, and {\em at a certain stage in our
development we will cease to be law-seekers and become
law-makers}.\footnote{For more discussion on this theme, see the
section in \cite{rap11}, titled `{\em The Saviors of Physical
Law}', emulating Kazantzakis' ``{\em The Saviors of God}''
\cite{kaz}. Our emphasis.}

It is not clear what happens to the concept of a correct theory
when we abandon the notion that it is a faithful picture of
nature. Presumably, just as the theory is an aspect of our
collective life, its truth is an aspect of the quality of our
life...}''
\end{quotation}

\noindent And what better means other than our Logos---or better,
than our imaginative and creative Logos: our poetic and bardic
{\em Mythos}---do we possess for approximating the archegonal
Truth about Nature? Moreover, what a humbling thought this is:
that in the end we may find out that this truth is the
quintessential quality of our ellogous lives. Then, in a
Nietzscheic sense \cite{nehamas}, {\em we will have become what we
already are: Poets true to our Nature!}

\end{document}